\documentclass[prd,eqsecnum,showpacs,nofootinbib,superscriptaddress]{revtex4}
\usepackage{bm}
\usepackage{stmaryrd}
\usepackage{amsmath}
\usepackage{amssymb}
\usepackage{graphicx}
\usepackage{textcomp}
\usepackage{calrsfs}
\usepackage{yfonts}

\newcommand{\nn}{\nonumber}

\newcommand{\be}{\begin{equation}}
\newcommand{\ee}{\end{equation}}
\newcommand{\ben}{\begin{equation*}}
\newcommand{\een}{\end{equation*}}
\newcommand{\bea}{\begin{eqnarray}}
\newcommand{\eea}{\end{eqnarray}}

\DeclareMathOperator{\Ai}{Ai}
\DeclareMathOperator{\Bi}{Bi}

\begin{document}
\title{Stress tensor for a scalar field
in a spatially varying  background potential: Divergences,
``renormalization,'' anomalies, and Casimir forces}
\author{Kimball A. Milton}\email{milton@nhn.ou.edu}
\affiliation{Homer L. Dodge Department of Physics and Astronomy,
University of Oklahoma, Norman, OK 73019-2061, USA}
\author{Stephen A.  Fulling}\email{fulling@math.tamu.edu}
\affiliation{Departments of Mathematics and Physics, Texas A\&M University,
College Station, TX 77843-3368, USA}
\author{Prachi Parashar}\email{prachi@nhn.ou.edu}
\affiliation{Department of Physics, Southern Illinois University-Carbondale,
Carbondale, IL 62091-6899, USA}
\affiliation{Homer L. Dodge Department of Physics and Astronomy,
University of Oklahoma, Norman, OK 73019-2061, USA}
\author{Pushpa Kalauni}\email{pushpakalauni60@gmail.com}
\affiliation{Homer L. Dodge Department of Physics and Astronomy,
University of Oklahoma, Norman, OK 73019-2061, USA}
\author{Taylor Murphy}\email{taylormurphy@ou.edu}
\affiliation{Homer L. Dodge Department of Physics and Astronomy,
University of Oklahoma, Norman, OK 73019-2061, USA}

\date\today

\begin{abstract}
Motivated by a desire to understand quantum fluctuation energy densities and
stress within a spatially varying dielectric medium, we examine the vacuum
expectation value for the stress tensor of a scalar field with arbitrary
conformal parameter, in the background of a given potential that depends
on only one spatial coordinate.  
We regulate the expressions by incorporating a temporal-spatial cutoff in
the (imaginary) time and transverse-spatial directions.  The divergences are
captured by the zeroth- and second-order WKB approximations.  Then the 
stress tensor is ``renormalized'' 
by omitting the terms that depend on the cutoff.
The ambiguities that inevitably arise in this procedure are both duly noted
and restricted by imposing certain physical conditions; one result is that
the renormalized stress tensor exhibits the expected trace anomaly.
The renormalized stress tensor exhibits no pressure anomaly,
in that the principle of virtual work is satisfied for motions in a transverse
direction. We then consider
a potential that defines a wall, a one-dimensional potential that vanishes
for $z<0$ and rises like $z^\alpha$, $\alpha>0$, for $z>0$.  Previously,
the stress tensor had been computed outside of the wall, whereas now we
compute all components of the stress tensor in the interior of the wall.
The full finite stress
tensor is computed numerically for the two cases where explicit solutions to
the differential equation are available, $\alpha=1$ and 2.  The energy density
exhibits an inverse linear divergence as the boundary is approached from the
inside for a linear potential, and a logarithmic divergence for a quadratic
potential.  Finally, the interaction between two such walls is computed,
and it is shown that the attractive Casimir pressure between the two walls
also satisfies the principle of virtual work (i.e., the pressure equals 
the negative derivative of the energy with respect to the distance between the
walls).   
\end{abstract}


\maketitle

\section{Introduction}
When Casimir discovered \cite{Casimir} that, because of quantum fluctuations, 
two uncharged 
perfectly conducting parallel plates attracted each other in vacuum, he
considered an unphysical abstraction.  Lifshitz \cite{Lifshitz} partially
remedied this defect, by allowing the plates to consist of dielectric
material with an arbitrary permittivity as a function of frequency, but he
still imagined that the plates were separated by vacuum.  This omission was
removed a few years later with the addition of Dzyaloshinskii and Pitaevskii
\cite{Dzyaloshinskii}; now the plates could be separated by another dielectric.
But the geometry still was a three-layer system: the dielectric material 
was spatially constant in each region.  The general problem of a spatially
varying medium has still not been solved \cite{Leonhardt}. (Recent papers
on this topic include
Refs.~\cite{Leonhardt:2011zz,Graham,Beauregard:2014joa}.)
It is not merely a matter of
numerics: Divergences arise associated with this variation that are still
not understood.  For an overview of the state of knowledge in Casimir physics,
see Ref.~\cite{Dalvit}.  In this paper we use natural units, with $\hbar=c=1$.

Some years ago we started a program to investigate such problems in the 
context of a simpler scalar field interacting with a spatially varying 
potential.  The proposal of a soft wall was made in Ref.~\cite{Bouas:2011ik};
that is, we consider a potential of the form 
\be
v(z)=\left\{\begin{array}{cc}
0,&z<0,\\
z^\alpha,&z>0,\end{array}\right.
\ee
with $\alpha>0$, the coupling
constant being absorbed into the definition of $z$ \cite{Bouas:2011ik}.
This potential interacts with a massless scalar field $\phi$ 
governed by the Lagrangian
\be
\mathcal{L}=-\frac12\partial_\mu\phi\partial^\mu\phi-\frac{v}2 \phi^2.
\ee
The corresponding stress-energy tensor in flat Minkowski space with
$g^{\mu\nu}=\mbox{diag}(-1,1,1,1)$ is
\be
T^{\mu\nu}=\partial^\mu\phi\partial^\nu\phi-\frac12 g^{\mu\nu}
\left(\partial_\lambda\phi\partial^\lambda\phi+v\phi^2\right)-\xi\left(
\partial^\mu\partial^\nu-g^{\mu\nu}\partial^2\right)\phi^2.
\label{stresstnensor}
\ee
Here, we have included the arbitrary conformal term, with the conformal 
parameter $\xi$.  The value $\xi=1/6$ is the one that makes conformal
symmetry manifest, and softens many divergences in scalar quantum field theory.

In Ref.~\cite{hs} we computed the energy density for this model, mostly
in the vacuum region below  the wall.  Once the bulk energy density,
which makes no reference to the potential at all, is subtracted,  the energy
density is finite outside the wall.  
We showed that the energy density diverges as the
boundary is approached from below,
\be
z\to 0^-:\quad u(z)\sim \frac{1-6\xi}{96\pi^2}\frac1z,\label{surfdivlin}
\ee
for a linear wall,
much softer than the $1/z^4$ divergence seen for a Dirichlet wall.
For a quadratic wall, the surface divergence is logarithmic,
\be
z\to 0^-:\quad u(z)\sim -\frac{1-6\xi}{48\pi^2}\Gamma(0,2|z|).
\label{surfdivquad}
\ee
For larger values of $\alpha$ there is no surface divergence at all.

We also analyzed the divergence structure within the walls, where using
second-order WKB analysis we showed with a temporal cutoff $\tau$ that
the energy density had the following dependence:
\be
u(z)\sim \frac3{2\pi^2}\frac1{\tau^4}-\frac1{8\pi^2}\frac{v(z)}{\tau^2}
+\frac1{32\pi^2}\left[v(z)^2+\frac23(1-6\xi)\frac{\partial^2}{\partial z^2}
v(z)\right]\ln\tau.\label{udiv}
\ee
The first term is the divergent bulk energy density, independent of the 
potential.  The lower-order divergences involve the potential.

The first steps in extending this work have been given in Ref.~\cite{murray}.
There, general formulas are given for all components of the stress tensor,
and a strategy for extending the computational ability to general $\alpha$
is sketched.  Here we tackle the general problem within the wall, but with
detailed numerical results restricted to the explicitly solvable cases
$\alpha=1,2$.  

In Sec.~\ref{sec:gf} we state the general Green's function formulation
of the problem, and discuss the point-splitting regulation scheme used to
define the vacuum expectation value of the square of the field. We then
give formulas for constructing the vacuum expectation value of the stress
tensor.  In Sec.~\ref{sec:asy} we identify the divergences occurring in the
vacuum expectation value of the stress tensor, based on the second-order
WKB approximation.
In Sec.~\ref{sec:trace} we give the classically expected trace and
divergence equations satisfied by the stress tensor.  Since the WKB solutions
found in Sec.~\ref{sec:asy} are only approximate, the divergence,
or conservation, identity
is only approximately satisfied in any order of WKB approximation, although
the trace identity is automatically satisfied for any $g$.

The divergences found in Sec.~\ref{sec:asy} are systematically discarded in
Sec.~\ref{sec:ren}.  As in the curved-space analogue, at least some of these
divergences correspond to terms in the original 
Lagrangian \cite{Mazzitelli:2011st,Bouas:2011ik,Fulling:2012zz}, 
so we shall refer to this process as ``renormalization.'' 
There are logarithmically divergent terms;
these transform into finite terms depending logarithmically on the potential 
with an arbitrary mass scale.  The process of renormalizing the stress tensor
involves two further steps: the vacuum expectation value of the scalar field 
is shifted by an amount proportional to the square of the cutoff parameter 
$\delta$; and the stress tensor is modified by the addition of a term
proportional to the second heat-kernel coefficient, so that it does not
possess a conservation anomaly.  As a consequence, the stress tensor acquires
a trace anomaly.  In this procedure we follow Wald \cite{wald}.  The resulting
renormalized stress tensor is now diagonal and satisfies the principle of 
virtual work, displaying  no pressure anomaly.

    The considerations in Secs.~\ref{sec:gf}-\ref{sec:ren} 
are more general than the 
steeply rising potential considered in the rest of the paper.  
They apply to (at least) any positive Klein-Gordon potential that 
depends on only one Cartesian coordinate.

We then go on in Secs.~\ref{sec:lin} and \ref{sec:quad} to discuss the energy
density in the interior region for the linear and the quadratic potentials,
respectively.  We compute the finite remainders numerically, and show
that they have the expected divergences as the boundary $z=0$ is approached 
from above, the same as those found outside (further discussion in Appendix
\ref{sec:surf}).  The behavior of $\langle T_{zz}\rangle$, which does not 
exhibit any surface divergence, is discussed in Sec.~\ref{sec:tzz}.

In the final section \ref{sec:2walls} we consider two such walls, with
arbitrary, mirrored, potentials.  The Lifshitz formula is easily obtained for
the force between the walls, which is shown, for arbitrary potential, to
be equally well derivable from the total energy obtained by integrating
the energy over the regions between as well as inside the potentials.  Thus,
as expected, the principle of virtual work is once again satisfied.

The Conclusion discusses further directions this work will pursue.  It is 
followed by two appendices, one on the WKB approximation and the second
on the derivation of the ``surface divergences.''

\section{Green's function and construction of stress tensor}
\label{sec:gf}
We will compute in this paper the vacuum expectation value
of the stress tensor 
 obtained from the Green's 
function, which for this $(2+1)$ dimensional spatial geometry has the form
\be
G(x,x')=\int\frac{d\omega}{2\pi}\frac{(d\mathbf{k}_\perp)}{(2\pi)^3}
e^{-i\omega(t-t')}e^{i\mathbf{k_\perp\cdot(r-r')_\perp}}g(z,z';\omega,
\mathbf{k}_\perp),
\ee
which satisfies the differential equation
\be
\left(\frac{\partial^2}{\partial t^2}-\nabla^2+v\right)G(x,x')=\delta(x-x'),
\ee
where $x^\mu$ is a four-vector, and so the delta-function is a four-dimensional
one.  The time-ordered product of fields is the quantum
correspondent of this Green's function,
\be
\langle \mbox{T} \phi(x)\phi(x')\rangle=\frac1iG(x,x').\label{vev}
\ee
It is more than convenient to perform a Euclidean transformation (more than
a Wick rotation)
\be
\omega\to i\zeta,\quad (t-t')\to i(\tau-\tau'),
\ee
which is permitted because the Green's function has no singularities in
the first and third quadrants.  Then the reduced Green's function becomes a
function of $\kappa=\sqrt{\zeta^2+k^2}$, $k=|\mathbf{k}_\perp|$:
\be
g(z,z';\omega,\mathbf{k}_\perp)\to g(z,z';\kappa).
\ee
The reduced Green's function then satisfies
\be
\left(-\frac{\partial^2}{\partial z^2}+\kappa^2+v(z)\right)g(z,z')
=\delta(z-z').
\ee
In general, there is no closed form solution to the homogeneous version of this
equation; therefore in the next section, we give the leading and 
next-to-leading WKB approximations to the solutions of this equation (which
capture the asymptotic behavior in any case), and compute the 
corresponding stress tensor components, obtained by applying a differential
operator to 
Eq.~(\ref{vev}).  These are divergent, so we regulate them by point-splitting
in time and transverse space:
\be
\tau-\tau'\to\tau\to0,\quad (\mathbf{r-r'})_\perp\to \bm{\rho}\to 0.
\ee
Everything is expressed in terms of scalar integrals involving $\delta$
\cite{murray},
\be
\delta=\sqrt{\tau^2+\rho^2}.
\ee
In particular, the vacuum expectation value of $\phi^2$ is given by
\be
\langle \phi^2(x)\rangle=\frac1i\int\frac{d\omega}{2\pi}
\frac{(d\mathbf{k}_\perp)}{(2\pi)^2}
e^{i\mathbf{k}\cdot\bm{\delta}}g(z,z;\omega,
k)=\frac1{2\pi^2}\int_0^\infty d\kappa\,\kappa^2 g(z,z;\kappa)\frac{\sin\kappa
\delta}{\kappa\delta}\equiv I[g(z)].\label{vevphi2}
\ee
Here, $g(z)=g(z,z)$.

The general expression for the reduced Green's function can be taken to be
\be
g(z,z')=\frac1w F(z_>)G(z_<)-\frac1w F(z)F(z')\frac{G(0)-G'(0)/\kappa}
{F(0)-F'(0)/\kappa},\label{gee}
\ee
where $F$ is a solution of the homogeneous equation,
\be
\left(-\frac{\partial^2}{\partial z^2}+\kappa^2+v(z)\right)\left\{\begin{array}
{c}F\\G\end{array}\right\}=0,\label{homosln}
\ee
that decays at positive infinity,
and $w$ is the Wronskian with a second independent solution $G$,
\be
w=F(z)G'(z)-G(z)F'(z),\label{wronskian}
\ee
which is independent of $z$. It is important to note that adding an arbitrary
multiple of $F$ to $G$ does not change the Green's function.

All components of the stress tensor can be computed from the Green's function,
more particularly in terms of the regulated 
vacuum expectation value of $\phi^2$ (\ref{vevphi2}).
For example, the energy density is given by \cite{murray} ($\beta=\xi-1/4$)
\begin{subequations}
\label{stform}
\be
u=\left(\frac{\partial^2}{\partial \tau^2}-\beta\frac{\partial^2}{\partial z^2}
\right)I[g(z)],\label{enden}
\ee
and the $xx$ and $yy$ components of the stress tensor are expressed as
\be
\langle T_{xx}\rangle=-\left(\frac{\partial^2}{\partial \rho_x^2}
-\beta\frac{\partial^2}{\partial z^2}\right) I[g(z)],\label{txx}
\ee
and
\be
\langle T_{yy}\rangle=-\left(\frac{\partial^2}{\partial \rho_y^2}
-\beta\frac{\partial^2}{\partial z^2}\right) I[g(z)],\label{tyy}
\ee
while the $zz$ component is written as
\be
\langle T_{zz}\rangle=\frac14\frac{\partial^2}{\partial z^2}I[g(z)]
-I[(\kappa^2+v(z))g(z)].
\label{tzz}
\ee
The off-diagonal terms are given by
\be
\langle T_{xy}\rangle
=\frac\partial{\partial\rho_x}\frac\partial{\partial\rho_y}
I[g(z)],\quad
\langle T_{0x}\rangle
=i\frac\partial{\partial\tau}\frac\partial{\partial\rho_x}
I[g(z)],\quad
\langle T_{0y}\rangle
=i\frac\partial{\partial\tau}\frac\partial{\partial\rho_y}
I[g(z)],
\ee
while
\be
\langle T_{0z}\rangle=\langle T_{xz}\rangle=\langle T_{yz}\rangle=0.
\ee
\end{subequations}

\section{Asymptotic Behavior}
\label{sec:asy}
Now  we wish to obtain a generalization of Eq.~(\ref{udiv}), which mirrors
divergences much earlier obtained in curved space 
\cite{Christensen:1976vb,Bunch:1978gb,adler}. 
 The large $\kappa$ behavior to the integrand in (\ref{vevphi2}) 
is dominated by that of the first term in Eq.~(\ref{gee}).
At coincident points, that term 
 is approximated by the second WKB approximation \cite{bo}
\be
\frac{F(z)G(z)}w\sim \tilde g(z,z)\equiv
\frac1{2\sqrt{\kappa^2+v(z)}}-\frac{v''(z)}{16(\kappa^2+v(z))^{5/2}}
+\frac{5 v^{\prime2}(z)}{64(\kappa^2+v(z))^{7/2}}.\label{2wkb}
\ee

    In Eqs.\ (\ref{I0}) and (\ref{I2}) we have expanded $I[\tilde{g}]$ 
through order $\delta^2$, thereby obtaining an approximation to 
$I[g]$ that is second-order in both WKB and point-splitting 
senses.  Inserting it into Eqs.\ (\ref{stform}), we obtain the stress 
tensor to second WKB order and to $O(\delta^0)$:
\bea
\langle\tilde T_{\mu\nu}\rangle
&=&\frac1{2\pi^2\delta^4}\left(\begin{array}{cccc}
\frac{3\tau^2-\rho^2}{\delta^2}&\frac{4i\tau\rho_x}{\delta^2}
&\frac{4i\tau\rho_y}{\delta^2}&0\\
\frac{4i\tau\rho_x}{\delta^2}&\frac{\tau^2+\rho_y^2-3\rho_x^2}{\delta^2}&
\frac{4\rho_x\rho_y}{\delta^2}&0\\
\frac{4i\tau\rho_y}{\delta^2}&\frac{4\rho_x\rho_y}{\delta^2}&
\frac{\tau^2+\rho_x^2-3\rho_y^2}{\delta^2}&0\\
0&0&0&1\end{array}\right)+\frac{v(z)}{8\pi^2\delta^2}
\left(\begin{array}{cccc}
\frac{\rho^2-\tau^2}{\delta^2}&-\frac{2i\tau\rho_x}{\delta^2}
&-\frac{2i\tau\rho_y}{\delta^2}&0\\
-\frac{2i\tau\rho_x}{\delta^2}&\frac{\rho_x^2-\tau^2-\rho_y^2}{\delta^2}&
-\frac{2\rho_x\rho_y}{\delta^2}&0\\
-\frac{2i\tau\rho_y}{\delta^2}&-\frac{2\rho_x\rho_y}{\delta^2}&
\frac{\rho_y^2-\rho_x^2-\tau^2}{\delta^2}&0\\
0&0&0&-1\end{array}\right)\nn\\
&&\quad\mbox{}+\frac{v^2(z)}{32\pi^2}\left(\ln\frac{\sqrt{v}\delta}2+\gamma
\right)\mbox{diag}(1,-1,-1,-1)-\left(\xi-\frac16\right)\frac{v''(z)}{8\pi^2}
\left(\ln\frac{\sqrt{v}\delta}2+\gamma
\right)\mbox{diag}(1,-1,-1,0)\nn\\
&&\quad\mbox{}+\frac{v^2(z)}{128\pi^2}
\left(\begin{array}{cccc}
\frac{\tau^2-3\rho^2}{\delta^2}&\frac{4i\tau\rho_x}{\delta^2}
&\frac{4i\tau\rho_y}{\delta^2}&0\\
\frac{4i\tau\rho_x}{\delta^2}&\frac{3\tau^2+3\rho_y^2-\rho_x^2}{\delta^2}&
\frac{4\rho_x\rho_y}{\delta^2}&0\\
\frac{4i\tau\rho_y}{\delta^2}&\frac{4\rho_x\rho_y}{\delta^2}&
\frac{3\tau^2+3\rho_x^2-\rho_y^2}{\delta^2}&0\\
0&0&0&3\end{array}\right)-\frac{v''(z)}{96\pi^2}\left(\begin{array}{cccc}
\frac{\tau^2}{\delta^2}&\frac{i\tau\rho_x}{\delta^2}
&\frac{i\tau\rho_y}{\delta^2}&0\\
\frac{i\tau\rho_x}{\delta^2}&-\frac{\rho_x^2}{\delta^2}&
\frac{\rho_x\rho_y}{\delta^2}&0\\
\frac{i\tau\rho_y}{\delta^2}&\frac{\rho_x\rho_y}{\delta^2}&
-\frac{\rho_y^2}{\delta^2}&0\\
0&0&0&0\end{array}\right)\nn\\
&&\quad\mbox{}+\frac1{96\pi^2}\left[6\frac{v^{\prime2}(z)}{v(z)}
-\frac{\partial^2}{\partial z^2}\left(\frac{v''(z)}{v(z)}\right)\right]
\mbox{diag}(-\beta,\beta,\beta,1/4)\nn\\
&&\quad\mbox{}+\frac1{384\pi^2}\left[\frac{v^{\prime2}(z)}{v(z)}
\mbox{diag}(-1,1,1,-5)
+2\frac{\partial^2}{\partial z^2}\left(\frac{v^{\prime2}(z)}{v^2(z)}\right)
\mbox{diag}(-\beta,\beta,\beta,1/4)\right].\label{leadingt}
\eea
(Here, the tilde notation means that the 2nd WKB approximation is being
employed.)
Of the ten terms displayed above, 
the last two give the finite contribution from the final term in 
Eq.~(\ref{2wkb}). The middle term in Eq.~(\ref{2wkb}), 
which also arises from the 2nd order
WKB approximation as discussed in Appendix \ref{sec:wkb}, contributes both
 to the 8th term in Eq.~(\ref{leadingt}) and to the
divergent and ambiguous terms proportional to $v''$, the 6th and 4th terms,
 and results in the
conversion of $\beta$ to the expected $\xi-1/6$ in the 4th term above.
The remaining terms arise from the 0th order WKB approximation, the first term
in Eq.~(\ref{2wkb}).

    It is obvious that (at least) the most singular terms in Eq.\
(\ref{leadingt}) can be written in a covariant tensorial form in analogy to 
the formulas of Christensen \cite{Christensen:1976vb} for the case of an 
external gravitational field.  We find it convenient, however, to 
delay displaying the result of that step until after 
renormalization [see Eq.~(\ref{rst})].

\section{Trace and Divergence Theorems}
\label{sec:trace}

From Eq.~(\ref{stresstnensor}) 
we can immediately show, classically, that the trace of the stress tensor is
\be
T^\mu{}_\mu=-v\phi^2+\frac12(6\xi-1)\partial^2\phi^2,\label{traceid}
\ee
while the divergence is
\be
\partial_\mu T^{\mu\nu}=-\frac12 \phi^2\partial^\nu v.\label{divid}
\ee
As expected, the stress tensor is conserved outside the potential region, and
is traceless there as well for conformal coupling, $\xi=1/6$.

What happens quantum-mechanically? Consider first the trace.  According to
Eqs.~(\ref{stform}) the vacuum expectation value of the trace is
\be \langle T^\mu{}_\mu\rangle=
-\langle T^{00}\rangle+\langle T_{xx}\rangle+\langle T_{yy}\rangle
+\langle T_{zz}\rangle=\left[-\left(\frac{\partial^2}{\partial\tau^2}
+\frac{\partial^2}{\partial \rho_x^2}+\frac{\partial^2}{\partial \rho_y^2}
\right)+\left(3\beta+\frac14\right)\frac{\partial^2}{\partial z^2}
\right]I[g]-I[(\kappa^2+v)g].
\ee
But $I[g]$ depends on the cutoff parameters  only through the combination
 $\delta$, so the above is simply
\be
\langle T^\mu{}_\mu\rangle=-\left[\frac{\partial^2}{\partial\delta^2}
+\frac2\delta\frac{\partial}{\partial
\delta}-3\left(\xi-\frac16\right)\frac{\partial^2}{\partial z^2}\right]I[g]
-I[(\kappa^2+v)g].\label{tr1}
\ee
Using
\be
\left(-\frac{d^2}{d\delta^2}-\frac2\delta\frac{d}{d\delta}\right)\frac{\sin
\kappa\delta}{\kappa\delta}=\kappa^2\frac{\sin\kappa\delta}{\kappa\delta},
\label{deltalap}
\ee
we simplify Eq.~(\ref{tr1}) to
\be
\langle T^\mu{}_\mu\rangle=-vI[g]+3\left(\xi-\frac16\right)\partial_z^2 I[g],
\label{qtraceid}
\ee
which is the vacuum expectation value of the classical trace identity 
(\ref{traceid}).  This is true identically as a functional relationship for 
any $g$, so it is satisfied exactly by the WKB approximation, to any order.

What about the divergence equation (conservation law)?  
The nonzero component of the divergence of the stress tensor is
\be 
\partial_\mu \langle T^{\mu z}\rangle=\partial_z\langle T^{zz}\rangle=
\frac14\partial_z^3 I[g]-\partial_z I[(\kappa^2+v)g],
\ee
and the question is whether this is equal to $-\frac{v'}2 I[g]$.
This will be an identity if $I$ is a functional of the exact Green's function 
which satisfies 
\be
(-\partial_z^2+\kappa^2+v)g(z,z')=0, \quad z\ne z'.
\ee
  But the WKB approximants
do not satisfy the equation of motion.  In fact, if we use the zeroth
order approximation given in Eq.~(\ref{leadingt})
(essentially the first 7 terms there), 
we find that (terms that vanish with $\delta$ are omitted here and in the
following)
\be
\partial_z\langle \tilde T_{zz}^{(0)}\rangle+\frac{v'}2I[g^{(0)}]
=\frac14\partial_z^3I[g^{(0)}]-\partial_zI[(\kappa^2+v)g^{(0)}]+\frac{v'}2
I[g^{(0)}]=\partial_z\frac1{64\pi^2}\left(\frac{v^{\prime2}}v\right),
\ee
where the right side is simply 
the $z$-derivative of the unambiguous finite
part of the stress tensor originating in this order, the first term in the
penultimate line of Eq.~(\ref{leadingt}), the 7th term.  Note that this
zeroth-order discrepancy is third-order in derivatives.
If we include both the zeroth and
second order terms (all the terms displayed in Eq.~(\ref{leadingt})) the
discrepancy is fifth-order in derivatives:
\be
\frac14\partial_z^3I[g^{(0)+(2)}]-\partial_zI[(\kappa^2+v)g^{(0)+(2)}]
+\frac{v'}2I[g^{(0)+(2)}]
=-\partial_z^3\frac1{384\pi^2}\left(\frac{v''}{v}-\frac{v^{\prime2}}{2v^2}
\right)=\frac14\partial_z^3I[g^{(2)}],\label{divan2}
\ee
where $I[g^{(2)}]$ is 
given in Eq.~(\ref{I2}).  [Alternatively, it is the $z$-derivative of the
8th and 10th terms in Eq.~(\ref{leadingt}).]
If we go through the fourth order, we get
\be
\frac14\partial_z^3 I[g^{(0)+(2)+(4)}]-\partial_zI[(\kappa^2+v)g^{(0)+(2)+(4)}]
+\frac{v'}2I[g^{(0)+(2)+(4)}]=\frac14\partial_z^3I[g^{(4)}],
\ee
where $I[g^{(4)}]$ is given in Eq.~(\ref{I4}).  
The discrepancy is now 7th-order in derivatives.
In each case the lower-order discrepancy is cancelled, and the
remaining discrepancy is pushed to the next-higher order.

\section{Renormalization}
\label{sec:ren}

    We now wish to obtain finite, ``renormalized'' values for 
$\langle \phi^2 \rangle$ and $\langle T_{\mu\nu} \rangle$.  The 
former is needed both to investigate the fate of the trace and 
divergence identities of Sec.~\ref{sec:trace} and to provide a simple way of 
getting the renormalized stress tensor itself in parallel to the 
derivation of the regularized version (\ref{leadingt}); for the latter 
purpose we need to keep the $O(\delta^2)$ terms in $I[g]$.

    Naively, one would like simply to discard from (\ref{leadingt}) all 
terms that, as $\delta\to0$, either diverge or depend on the 
direction of the point-splitting vector $(\tau,\bm{\rho})$.  
The problem of justifying that step physically by a genuine 
renormalization of coupling constants in a full theory including 
the gravitational field and the scalar field $v$ as dynamical 
objects will not be discussed here (but see 
Refs.~\cite{Mazzitelli:2011st,Bouas:2011ik,Fulling:2012zz}), 
hence our use of quotation 
marks around ``renormalized''\negthinspace.  Another problem, however, 
cannot be postponed: It is impossible to separate logarithmically 
divergent terms from finite terms in a scale-invariant manner, 
and likewise it is impossible to separate direction-dependent 
terms from direction-independent finite terms unambiguously.  
Both of these ambiguities afflict only the terms proportional to 
$v^2$ and $v''$; we shall refer to all such terms as being of 
``critical order''\negthinspace.  What is being confronted here 
is a close analogue of the situation in quantum field theory in 
curved space-time that was resolved by intensive work in the late 
1970s \cite{christensen,wald,AdlerLieberman1978}
(see also Ref.~\cite{Moretti:1998rs} and related papers), and we follow 
those references rather closely.  The basic doctrine is that 
terms of critical order in the renormalized stress tensor are 
inherently ambiguous but can be constrained by certain physical 
requirements of conservation and covariance (tensoriality).

    We start with the expression for $I[\tilde g]\equiv 
I[g^{(0)}]+I[g^{(2)}]$.  Starting from Eqs. (\ref{I0}) and (\ref{I2}), we 
omit the $O(\delta^{-2})$ term and make the replacement
\be
\ln\frac{\sqrt{v}\delta}2+\gamma-\frac12\to \ln \frac{\sqrt{v}}{\mu},
\label{log}
\ee
where $\mu$ is some arbitrary mass scale, in every logarithmic 
term.  Note that the constant term on the left-hand side of rule 
(\ref{log}) is an arbitrary convention, since a change in it can be 
regarded as a redefinition of $\mu$, but it is important to adopt 
the \emph{same} convention in every instance.  The result is a 
tentatively renormalized expression
\be
4\pi^2\bar{I}_R[g^{(0)+(2)}]
=\left[\frac{v}2+\frac{\delta^2}{16}\left(v^2-\frac13v''
\right)\right]\ln\frac{\sqrt{v}}{\mu}-\frac1{24}\left(\frac{v''}v-\frac12
\frac{v^{\prime2}}{v^2}
\right)-\frac{\delta^2}{64}\left(3v^2+\frac13 \frac{v^{\prime2}}{v}\right).
\label{2ndordvev}
\ee
in which the term $-\frac{3}{64}\delta^2 v^2$ arises because one 
logarithm in Eq.~(\ref{I0}) has $-\frac54$ instead of $-\frac12$.

    If Eq.~(\ref{2ndordvev}) is inserted into Eqs.~(\ref{stform}) 
in place of the unrenormalized $4\pi^2 I[g]$, one obtains a tentatively 
renormalized version of Eq.~(\ref{leadingt}) from which all divergent or 
direction-dependent terms have disappeared.  This calculation is 
facilitated by recognizing from Eq.~(\ref{deltalap}) that 
Eq.~(\ref{tzz}) may be replaced by
\be
\langle T_{zz}\rangle=\left[\frac14\partial_z^2+\nabla_\delta^2-v\right]I[g],
\ee
where we see the appearance of the Laplacian in the $\bm{\delta}$ coordinates,
which on a spherically symmetric function becomes
\be
\nabla^2_\delta=\frac{\partial^2}{\partial\delta^2}+\frac2\delta\frac\partial
{\partial\delta}.
\ee
In particular, $\nabla_\delta^2 \delta^2 = 6$.  However, it 
remains to grapple with the arbitrariness in the terms of 
critical order produced by this tentative procedure.

Adler et al.~\cite{adler} and Wald \cite{wald} demanded that (in 
our terminology) the terms subtracted from $I[\tilde g]$ to yield 
$I_R[\tilde g]$ must themselves be the leading asymptotic terms 
of a certain minimal solution of the Green's function's 
differential equation.  For technical reasons we find it hard to 
follow Wald's procedure in our setting, but we offer a different 
physical argument that leads in the end to the same result, in 
the sense that our trace anomaly (\ref{tranomaly}) agrees with Wald's 
general formula.  We observe that all terms of critical order in 
the tentatively renormalized $\langle T_{\mu\nu}\rangle$ can be 
written tensorially in terms of the metric tensor and $v$ and its 
covariant derivatives, with one exception, traceable to the term 
$-\frac{3}{64}\delta^2 v^2$ in $\bar{I}_R$ previously 
noted.  The bad term in the stress tensor can be removed by 
modifying the critical-order terms in Eq. (\ref{2ndordvev}): $I_R \equiv 
\bar I_R + \Delta I_R$ with
\be
4\pi^2\Delta I_R\equiv\frac3{64}\delta^2\left(v^2-\frac13 v''\right).
\ee
Here we see the appearance of 
\be
a_2=\frac12\left(v^2-\frac13 v''\right),
\ee
the second heat-kernel coefficient for the system under study
\cite[Sec.\ 4.8]{gilkey}, \cite[Chap.\ 9]{fulling},
which also occurs in the logarithmic term in Eq.~(\ref{2ndordvev}).
This gives a modification, for
example, to the $zz$ component of the stress tensor,
\be
4\pi^2\Delta \langle T_{zz}\rangle=\frac9{16}a_2.
\ee

We now follow Wald \cite{wald} precisely, observing that the 
critical-order terms in the new renormalized stress tensor do not 
obey the conservation law (\ref{divid}):
\be
4\pi^2\left[\partial_z \langle T_{zz}\rangle[I_R]+\frac12 v'I_R\right]
=-\partial_z \frac1{16} a_2.
\label{divanomaly}
\ee
(This phenomenon is entirely separate from the WKB residual 
indicated in Eq.~(\ref{divan2}), which involves terms of higher order in 
derivatives and does not represent any anomaly in the exact 
stress tensor.)  This ``conservation anomaly'' is cured by adding 
to the stress tensor another critical-order term:
\be
\langle T^{\mu\nu}\rangle_R=\langle T^{\mu\nu}\rangle[I_R]
+\frac{a_2}{64\pi^2}g^{\mu\nu}.\label{tmunuren}
\ee
This step introduces a trace anomaly,
\be
\langle T^\mu{}_{\mu}\rangle_R+vI_R-3\left(\xi-\frac16\right)\partial_z^2I_R
=\frac1{16\pi^2}a_2.\label{tranomaly}
\ee

So with this set of redefinitions, we are finally led to the following
form of the renormalized energy-momentum tensor:
\bea
4\pi^2\langle T^{\mu\nu}\rangle_R(z)
&=&-g^{\mu\nu}\frac{v^2}8\ln\frac{\sqrt{v}}{\mu}-
\frac12\left(\beta+\frac1{12}\right)(\partial^\mu\partial^\nu-
g^{\mu\nu}\partial^2)\left(v\ln\frac{\sqrt{v}}\mu\right)
+g^{\mu\nu}\frac{v^2}{32}\nn\\
&&\quad\mbox{}+\frac1{96}\frac{v^{\prime2}}v\mbox{diag}(1,-1,-1,1)
-\frac1{24}\partial_z^2\left(\frac{v''}v-\frac12\frac{v^{\prime2}}{v^2}\right)
\mbox{diag}\left(-\beta,\beta,\beta,\frac14\right)+4\pi^2t^{\mu\nu},
\label{rst}
\eea
where $t^{\mu\nu}$ is the remainder of the stress tensor, obtained by the
construction in Eqs.~(\ref{stform})  when the first and second WKB approximations
are subtracted from the Green's function,
\be
 t^{\mu\nu}=\langle T^{\mu\nu}\rangle-\langle \tilde T^{\mu\nu}\rangle=
\langle T^{\mu\nu}\rangle[I[g-\tilde g]].\ee
The form of the renormalized stress tensor 
(\ref{rst}) is a central result of this paper.

    Note that the terms of critical order [in the top line of 
Eq.~(\ref{rst})] are now completely tensorial.  The terms in the 
second line are, strictly speaking, part of the finite remainder, 
which need not be covariant in that sense.  Note also that the 
off-diagonal tensor components, which in Eq.~(\ref{leadingt}) were entirely 
direction-dependent, have now completely disappeared.  The 
direction-independent off-diagonal finite
terms must vanish by reflection symmetry in 
each of the coordinates $t$, $x$, and $y$. 
The renormalized $\langle T_{xx}\rangle_R$ and $\langle T_{yy}\rangle_R$ 
are the same as $u_R=\langle T^{00}\rangle_R$, except for a reversal of sign,
see Eqs.~(\ref{txx}) and (\ref{tyy}).  This proves
the nonexistence of a transverse pressure anomaly \cite{Fulling:2012zz}, 
completing the argument in Ref.~\cite{murray}.  

We will compute $t^{\mu\nu}$ 
numerically in the following two sections, for a linear and a quadratic
potential, respectively, where explicit formulas for the exact 
Green's functions can be given.

\section{Energy Density for the Linear Wall}
\label{sec:lin}
Let's consider the energy density for the linear wall, within the region of
the potential, 
\be
z>0:\quad v(z)=z.
\ee
In this case, the renormalized WKB stess tensor (\ref{rst}) gives the 
leading contribution:
\be
\tilde u_R=\langle \tilde T^{00}\rangle_R
=\frac1{32\pi^2}z^2\left(\ln\frac{\sqrt{z}}{\mu}-\frac14\right)
-\frac\beta{16\pi^2 z}
-\frac1{384\pi^2}\left[\frac1z+\frac{12\beta}{z^4}\right].\label{wkb}
\ee
The remainder of the energy density comes from substituting $g-\tilde g$
into Eq.~(\ref{vevphi2}) and using the construction for the energy density
in terms of this vacuum-expectation value, Eq.~(\ref{enden}).
Since the integral defining the remainder
is convergent without the cutoff, we expand the cutoff 
factor, $\sin\kappa\delta/(\kappa\delta)=1-(\kappa\delta)^2/6+\dots$, 
and obtain the remainder
\be
(u-\tilde u)(z)=-\frac1{2\pi^2}\int_0^\infty d\kappa\,\kappa^2\left(\frac13
\kappa^2+\beta\frac{\partial^2}{\partial z^2}\right)\left[g(z,z)-\tilde g(z,z)
\right].\label{rem}
\ee
Here, the explicit Green's function for the linear potential is
\be
g(z,z)=\pi \Ai(\kappa^2+z)\Bi(\kappa^2+z)-\frac{(\kappa \Bi-
\Bi')(\kappa^2)}{(\kappa\Ai-\Ai')(\kappa^2)}\pi
\Ai^2(\kappa^2+z).\label{lineargee}
\ee

\begin{figure}
\centering
\includegraphics{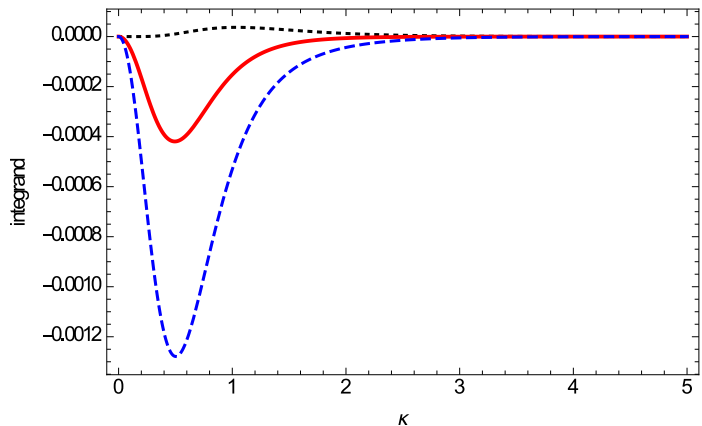}
\caption{\label{fig1} (color online)
Integrand in Eq.~(\ref{rem}) for $z=1$ for $\beta=0$ (dotted), $\beta
=-1/12$, the conformal value (thick), and $\beta=-1/4$ (dashed).
(All figures were prepared with \emph{Mathematica}.)}
\end{figure}

\begin{figure}
\centering
\includegraphics{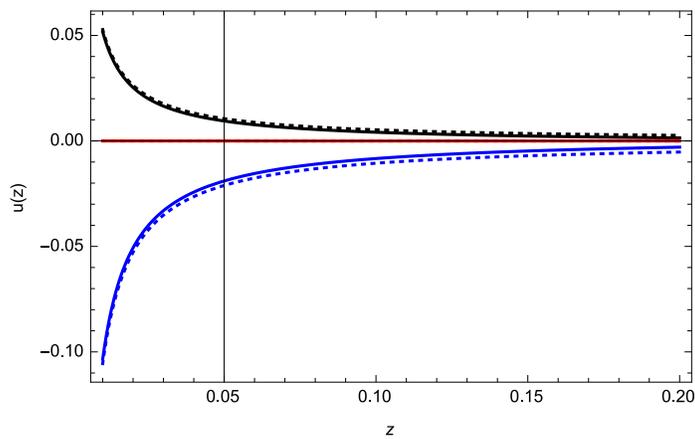}
\caption{\label{ulinint} 
(color online) Residual energy density within the wall for a linear
potential.  The solid curves show, from top to bottom, the energy density
for $\beta=0, -1/12, -1/4$, compared with the dotted curves which show the
surface energy estimates from Eq.~(\ref{usurf}).}
\end{figure}

We know from Ref.~\cite{murray}
that the WKB approximation is quite good for large
$\kappa$, so the integral (\ref{rem}) should converge quite rapidly.  The 
integrand is plotted in Fig.~\ref{fig1}.
To the numerical integration of the remainder (\ref{rem}) we add the portion
of the renormalized WKB energy from Eq.~(\ref{wkb}) that dominates for 
small values of $z$,
\be
u_{\rm wkb}=-\frac1{384\pi^2}\left(\frac{1+24\beta}z
+\frac{12\beta}{z^4}\right).
\label{uwkb}\ee
(The dominant terms in the renormalized energy density for large distances,
\be
u_{\rm leading}=
\frac1{32\pi^2}z^2\left(\ln\frac{\sqrt{z}}\mu-\frac14\right),\label{ua}
\ee
are ambiguous because they depend on the arbitrary scale $\mu$,
and require further discussion.)
In Fig.~\ref{ulinint} we plot the sum $u-\tilde u+u_{\rm wkb}$, which we
call the ``residual energy density.''
It is seen that that in each case the residual 
energy density rapidly goes to zero as
$z\to\infty$. (Of course, the terms in $u_{\rm leading}$, 
Eq.~(\ref{ua}), grow for large
$z$.) For the conformal case, $\beta=-1/12$, the energy density is nearly
zero.  Otherwise, the energy diverges as the boundary $z=0$ is
approached. For $\beta<-1/12$ the residual energy is everywhere negative, 
while for $\beta>-1/12$ the energy density is positive. 
The leading WKB divergence seen in 
Eq.~(\ref{uwkb}), going like $z^{-4}$, is clearly spurious, being canceled
by the remainder energy density (\ref{rem}). Thus, the dominant
structure in the integrand seen in Fig.~\ref{fig1} reflects this ultimately
spurious behavior, and the integrations have to be carried out to much
higher values of $\kappa$ than Fig.~\ref{fig1} suggests.
 The remaining divergence
near the boundary is 
precisely the same (in terms of $|z|$) as found in Ref.~\cite{hs} 
for the exterior region [see our Eq.~(\ref{surfdivlin})] 
\be
u_{\rm surf}=\frac{1+12\beta}{192\pi^2z}.\label{usurf}
\ee
This is shown as the dotted curves in Fig.~\ref{ulinint}.  In Appendix 
\ref{sec:surf} we give a plausibility argument for why this result might have
been expected.

\section{Energy Density for the Quadratic Wall}
\label{sec:quad}
\begin{figure}
\centering
\includegraphics{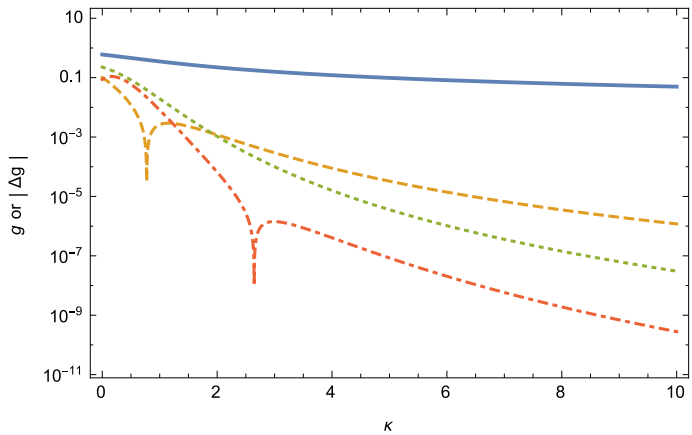} 
\caption{\label{gfcomp} (color online)
The reduced diagonal Green's function for the quadratic
wall, for $z=1$, shown by the solid curve.  The dashed curve shows the 
absolute value of the residual Green's function
after the leading WKB approximation is removed, the dotted curve shows the
residual after the two leading WKB approximations are subtracted, and the
dot-dashed curve shows the absolute value of the 
residual after the first three WKB approximations shown in Eq.~(\ref{quadrm})
are subtracted (top to bottom on the right).  Thus
the bottom curve shows the effect of subtracting the 
WKB approximations through second order. 
(Because what is plotted is the logarithm of the
absolute value of the residual of $g(z,z)$, the spikes occur at the points 
where the differences change sign.)}
\end{figure}

Now we are looking at 
\be
v(z)=z^2,\quad\alpha=2,
\ee
for which the renormalized second-order WKB energy density is
\be
\tilde u_R=\langle \tilde T^{00}\rangle_R
=\frac{z^4}{32\pi^2}\left(\ln\frac{z}\mu-\frac14\right)
-\frac{\xi-1/6}{4\pi^2}\ln\frac{z}\mu
-\frac1{48\pi^2}(1+18\beta).\label{wkbuq}
\ee
The remainder of the energy density is given by
\be
(u-\tilde u)(z)=-\frac1{2\pi^2}\int_0^\infty 
d\kappa\,\kappa^2\left(\frac{\kappa^2}3
+\beta\frac{\partial^2}{\partial z^2}\right)\left[g(z,z)-\frac1{2\sqrt{\kappa^2
+z^2}}+\frac1{8(\kappa^2+z^2)^{5/2}}
-\frac5{16}\frac{z^2}{(\kappa^2+z^2)^{7/2}}\right].\label{quadrm}
\ee
The diagonal Green's function for the quadratic wall is
\be
g(z,z)=\frac{\Gamma\left(\frac{\kappa^2+1}4\right)\Gamma\left(\frac{\kappa^2+3}
2\right)}{\pi 2^{(3-\kappa^2)/2}}\left[D_{-\frac{\kappa^2+1}2}(\sqrt{2}z)
D_{-\frac{\kappa^2+1}2}(-\sqrt{2}z)-\frac{\Gamma\left(\frac{\kappa^2+1}4\right)
-\frac2\kappa\Gamma\left(\frac{\kappa^2+3}4\right)}
{\Gamma\left(\frac{\kappa^2+1}4\right)
+\frac2\kappa\Gamma\left(\frac{\kappa^2+3}4\right)}
D^2_{-\frac{\kappa^2+1}2}(\sqrt{2}z)\right],
\ee
in terms of parabolic cylinder functions.
The WKB approximation is
very accurate, as shown in Fig.~\ref{gfcomp}.

\begin{figure}
\centering
\includegraphics{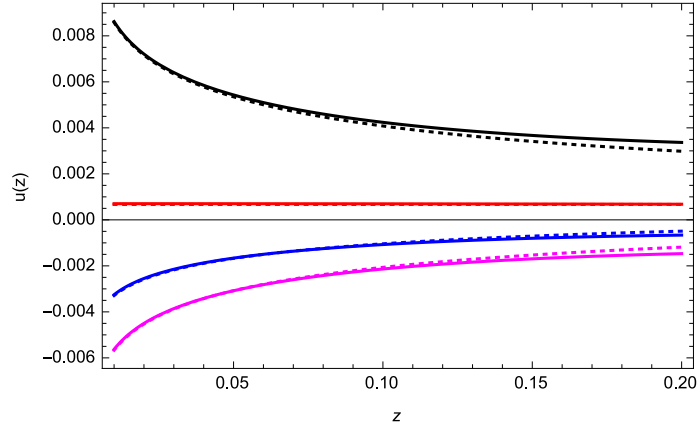}
\caption{\label{quaden} (color online)
Numerical integration of the residual energy [the sum of Eqs.~(\ref{quadrm})
plus (\ref{wkbquad})] for the quadratic wall
 for $\beta=1/20$, $\beta=0$, $\beta=-1/12$, the conformal 
value,  and $\beta=-1/4$, from bottom to top.  The results are insensitive
to the value of $K$, as long as it is sufficiently large, but not so large
that the errors in computing parabolic cylinder functions 
are significant. Here we used $K=10$. Here $\mu$ is
arbitrarily taken to be 1.  The dotted curves are the surface energies 
(\ref{surfquad}) with the offset (\ref{quadoff}).} 
\end{figure}

This time we add to the remainder energy, computed numerically, 
the parts of the renormalized WKB energy (\ref{wkbuq}) important for small $z$,
\be
u_{\rm wkb}=-\frac{1+12\beta}{48\pi^2}\ln\frac{z}
\mu-\frac1{48\pi^2}(1+18\beta).
\label{wkbquad}
\ee
(Again, we omit the leading term in the renormalized WKB energy
\be u_{\rm leading}=\frac{z^4}{32\pi^2}\left(\ln\frac{z}\mu-\frac14\right)
\ee
which is dominant for large distances.)
Convergence of the integral for the remainder energy is quite slow in this
case, complicated by the fact that \emph{Mathematica} 
fails to compute the parabolic
cylinder function accurately for large $\kappa$.  Therefore, it is necessary
to break up the integration into two parts,
\be
u(z)=\int_0^K d\kappa \mathcal{I}(\kappa,z)
+\int_K^\infty d\kappa \mathcal{I}(\kappa,z),\label{twoparts}
\ee
where $\mathcal{I}$ is the integrand shown in Eq.~(\ref{quadrm}), and then
compute the second integral, for large $K$, from the dominant WKB approximation
coming from the second term in Eq.~(\ref{gee}) [see Appendix \ref{sec:surf}, 
Eq.~(\ref{gf2q})]
\be
g_{F^2}\sim\frac1{16\kappa^5}e^{-2\kappa z},\quad \kappa\to\infty,\label{gf2}
\ee
which leads to the approximate evaluation
\be
\mathcal{I}(\kappa,z)\sim-\frac{1+12\beta}{96\pi^2}\frac1\kappa e^{-2\kappa z},
\quad\kappa\to\infty, \ee
and then to a form for the energy density suitable for numerical calculation,
\be 
(u-\tilde u)(z)\approx 
\int_0^K d\kappa\mathcal{I}(\kappa,z)-\frac{1+12\beta}{96\pi^2}
\Gamma(0, 2 K z),\label{largeK}
\ee
in terms of the incomplete gamma function.
Figure \ref{quaden} shows the residual energy density composed of the
remainder energy (\ref{quadrm}) [computed using Eq.~(\ref{largeK})] 
plus $u_{\rm wkb}$ [from Eq.~(\ref{wkbquad})]
for $\beta=0$, $-1/12$ (the conformal value),  $-1/4$, and $1/20$.
The method of Appendix \ref{sec:surf} yields a surface term
\be
u_{\rm surf}=-\frac{1+12\beta}{96\pi^2}\Gamma(0,2z),\label{surfquad}
\ee
which is, in fact, the same as the exterior result,
Eq.~(\ref{surfdivquad}), except, this time, for sign. However, a
constant term is undetermined by our asymptotic analyses.
To match the data, the surface energy is shifted by a constant amount,
which is empirically fitted by the simple formula
\be
u_{\rm offset}=0.00025(1-20\beta).\label{quadoff}
\ee
(The value for $\beta=1/20$ is shown in the graph to demonstrate that no
offset is required in that case.)  The fit is quite remarkably good.
The numerical fit, for large $\kappa$ and small $z$, is the statement
\be
\int_0^K d\kappa\mathcal{I}(\kappa,z)-\frac{1+12\beta}{48\pi^2}\ln z-\frac{1
+18\beta}{48\pi^2}+\frac{1+12\beta}{96\pi^2}\left[\Gamma(0,2z)-\Gamma(0,2Kz)
\right]\approx 0.00025(1-20\beta).\label{quadfit}
\ee
\section{Other stress tensor components}
\label{sec:tzz}

As noted above, $\langle T^{\mu\nu}\rangle_R$ is diagonal, and 
\be
\langle T_{xx}\rangle_R=\langle T_{yy}\rangle_R=-u_R.
\ee
So we only have to examine $\langle T_{zz}\rangle_R$.

\begin{figure}
\centering
\includegraphics{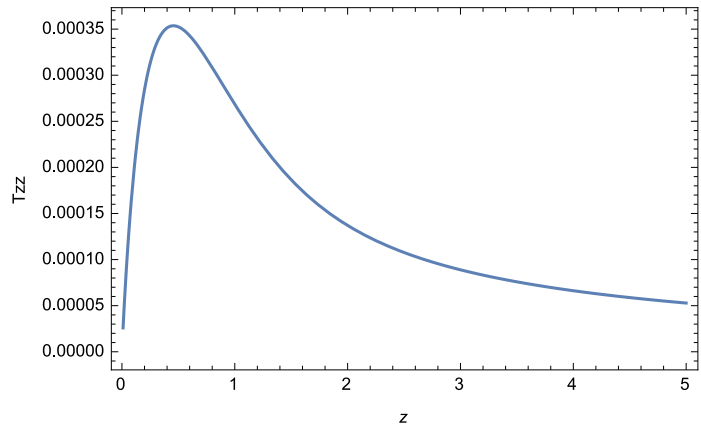}
\caption{\label{fig-tzz-lin}
This shows the  numerical values of remainder (\ref{tzzremainder}) added
to the WKB approximation (\ref{tzzwkb}) for the $zz$ component of the
stress tensor within the wall, for a linear potential.  The surface divergences
apparent in the WKB approximation are completely cancelled out by the numerical
remainder, leaving only a small residual.}
\end{figure}

For the linear wall
the 2nd-order renormalized WKB stress tensor (\ref{rst}) 
gives for the linear potential (omitting the ambiguous leading term
\be
T_{zz}^{\rm leading}=-\frac{z^2}{32\pi^2}\left(\ln\frac{\sqrt{z}}\mu-\frac14
\right),\label{leadtzz}
\ee
which is irrelevant for small $z$)
\be
\langle T_{zz}\rangle_{\rm wkb}=\frac1{384\pi^2}\left(\frac1z+\frac3{z^4}
\right).\label{tzzwkb}
\ee
This is added to the numerical evaluation of the remainder,
\be
t_{zz}=
\langle T_{zz}\rangle-\langle \tilde T_{zz}\rangle=\frac1{2\pi^2}\int_0^\infty
d\kappa\,\kappa^2\left(\frac14\frac{\partial^2}{\partial z^2}-(\kappa^2+z)
\right)\left[g(z,z)-\frac1{2\sqrt{\kappa^2+z}}-\frac5{64(\kappa^2+z)^{7/2}}
\right],\label{tzzremainder}
\ee
and 
$t_{zz}+\langle T_{zz}\rangle_{\rm wkb}$ is shown in Fig.~\ref{fig-tzz-lin}.
Now there is no surface divergence, and the residual stress tensor 
within the wall
[leaving aside the contribution of Eq.~(\ref{leadtzz})] is very small.
(The corresponding stress tensor outside the wall is exactly zero
\cite{hs,murray}.)
 
As for the trace and divergence identities, Eqs.~(\ref{tranomaly}) and 
(\ref{divanomaly}), these are not modified by the residual finite 
contributions.
The former is structurally true as noted in Eq.~(\ref{qtraceid}).  The latter
should be respected because the modification of the stress tensor 
(\ref{tmunuren})  involved only the critical terms.  
We have checked numerically
that our approximations are consistent:
\be
4\pi^2\left[\partial_zt_{zz}+\frac{v'}2(I-\tilde I)\right]
-\frac1{96}\partial_z^3
\left(\frac{v''}v-\frac{v^{\prime2}}{2v^2}\right)\to
2\int_0^\infty d\kappa\,\kappa^2\left[\frac14\partial_z^3
-(\kappa^2+z)\partial_z-\frac12\right](g-\tilde g)-\frac1{8z^5}\approx0,
\ee
consistent with zero within machine precision, 
where the substitution is for the linear wall.  Numerical consistency 
of the remainder stress tensor with the divergence identity has been 
verified also for the quadratic wall.

\section{Interaction between two mirrored soft walls}
\label{sec:2walls}
\begin{figure}
\includegraphics{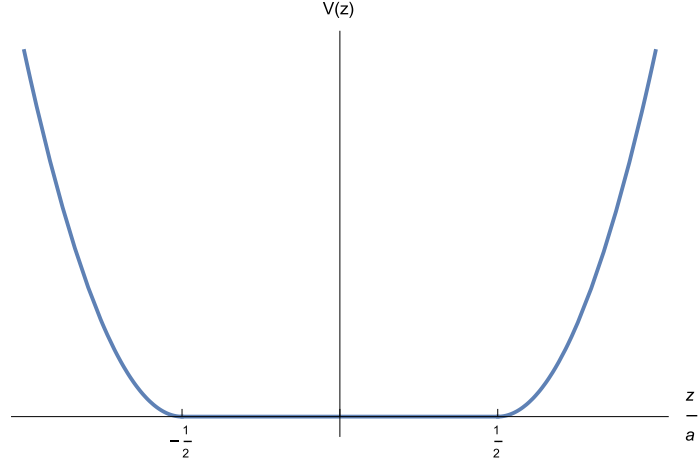}
\caption{\label{intwalls} Two facing soft walls, each modeled by a potential
$v(z)$, separated by a distance $a$.  The origin is chosen at the mid-point
between the two facing potentials.}
\end{figure}

Now imagine we have two such soft walls separated by a distance $a$, as
shown in Fig.~\ref{intwalls}.
 That is, let the potential be
\be
V(z)=\left\{\begin{array}{cc}
v(-z-a/2),&z<-a/2,\\
0,&-a/2<z<a/2,\\
v(z-a/2),&a/2<z.
\end{array}\right.
\ee
There is a Casimir force between the walls, and because each wall 
can move without changing its shape, it should be possible to 
calculate this force without depending upon the renormalization 
theory developed in the previous sections for the region inside 
the potential.  This indeed turns out to be the case.

    The reduced Green's function satisfies
\be
\left(-\frac{\partial^2}{\partial z^2}+\kappa^2+V(z)\right)g(z,z')
=\delta(z-z'),
\ee
which has the following solution in the three regions:
\begin{subequations}
\label{generalgf}
\bea
z<-a/2:&&\quad g(z,z')=\frac1w F(-z_<-a/2)G(-z_>-a/2)+R\frac1w F(-z-a/2)
F(-z'-a/2),\label{8.3a}\\
-a/2<z<a/2:&&\quad g(z,z')=\frac1{2\kappa}e^{-\kappa|z-z'|}+\frac1{2\kappa}
\frac{2r}{e^{2\kappa a}-r^2}\left[r\cosh\kappa(z-z') +e^{\kappa a}\cosh\kappa
(z+z')\right],\label{8.3b}\\
z>a/2:&&\quad g(z,z')=\frac1w F(z_>-a/2)G(z_<-a/2)+R\frac1w F(z-a/2)
F(z'-a/2).\label{8.3c}
\eea
\end{subequations}
Here $F$ and $G$ are independent solutions of the single potential 
problem (\ref{homosln}), with, again,
$F$ being the solution that vanishes at $z=+\infty$. 
Because the potential defines a cavity, in this section,
we will refer to the solutions $F$, $G$ as the ``exterior'' solutions, 
while the exponential solutions within the cavity are referred to as
``interior.'' 
 The Wronskian of the
two exterior solutions is $w$, Eq.~(\ref{wronskian}), 
is independent of $z$, and subsumes any normalization condition.
The reflection coefficients here
are generically computed by multiple scattering.  In terms of the abbreviations
\be
F_\pm=\kappa F(0)\pm F'(0),\quad G_\pm=\kappa G(0)\pm G'(0),\label{abbrev}
\ee
the interior (within the cavity) reflection coefficient is
\be
r=\frac{F_+}{F_-}<1, 
\ee
as already seen in Ref.~\cite{hs}.

The exterior (outside the cavity)
reflection coefficient $R$ is composed of the single-wall
exterior reflection coefficient $\bar r$,
\be
\bar r=-\frac{G_-}{F_-},
\ee
also as given in Ref.~\cite{hs}, followed by multiple reflections between the
interior walls,
\be
R=\bar r+\frac{r t^2}{e^{2\kappa a}-r^2},
\ee 
which involves the transmission coefficient across the wall (the same in 
either direction),
\be
t=\frac{\sqrt{2\kappa w}}{F_-},\label{tee}
\ee
where the numerator refers to the Wronskian $w$ for the exterior solutions
($F,G$) and the Wronskian ($2\kappa$) for the interior solutions 
($e^{\pm\kappa z}$).
Under an arbitrary scaling of the solutions, $F\to m F$, $G\to n G$,
where $m$ and $n$ are constants,
the Wronskian changes by $w\to mn w$, the interior reflection 
coefficient does not change, $r\to r$, while the exterior reflection 
coefficient and transmission coefficient change, 
\be \bar r\to\frac{n}m \bar r,\quad t\to\sqrt{\frac{n}m}t,
\ee
and hence the total reflection coefficient $R$ changes in the same way as
$\bar r$: $R\to \frac{n}m R$,  thus verifying the scaling consistency
of Eq.~(\ref{tee}).

Using only the interior reflection coefficient $r$ it is easy to calculate
the $zz$ component of the stress tensor in the vacuum region between the 
potentials, $-a/2<z<a/2$, using the prescription (\ref{tzz}).
The (divergent) contribution from the first term in Eq. (\ref{8.3b}), 
$1/(2\kappa)e^{-\kappa|z-z'|}$, is recognized as the universal 
zero-point pressure and discarded.  (This corresponds to the $T_{zz}$ 
component of the first term in Eq.~(\ref{leadingt}).)  The remainder leads 
immediately to the Lifshitz formula \cite{Lifshitz},
\be
t_{zz}=P=-\frac1{2\pi^2}\int_0^\infty d\kappa\,\kappa^3
\frac1{r^{-2}e^{2\kappa a}-1},\label{lifshitz}
\ee
which is independent of where it is evaluated in the cavity.  

If there is no additional pressure exerted on the system from 
infinity, $P$ is pressure felt by each wall; 
it is attractive, as expected.  In earlier sections, however, we 
have found terms in the renormalized stress tensor that grow as 
$|z|\to\infty$; the interpretation of these presumably unphysical 
terms is a topic for future work.

    We now wish to verify the principle of virtual work in the 
longitudinal direction, that is, that this pressure 
is the negative derivative of the total energy of the system with respect to 
the distance between the walls,
\be
P=-\frac{\partial U}{\partial a},\label{pvw}
\ee
where $U$ is the integral of the energy density over the entire system, the
energy per unit area,
\be
U=\int_{-\infty}^\infty dz\, u(z),
\ee
where $u$ is obtained by the operations given in Eq.~(\ref{enden}). 
The term proportional to $\beta$ vanishes because it is a total
derivative; there is no dependence on the conformal parameter
in the total energy.  Then, temporarily ignoring the first terms in each 
of Eqs.~(\ref{generalgf}), we 
calculate
\be
U-U_0=-\frac1{6\pi^2}\int_0^\infty d\kappa\,\kappa^4\left[\frac{2R}w
\int_0^\infty dz\, F^2(z)+\frac1{2\kappa}\frac{r}{e^{2\kappa a}-r^2}
\left(2ra+\frac1\kappa\left(e^{2\kappa a}-1\right)\right)\right].\label{U}
\ee
The contribution $U_0$ of the ignored terms is divergent but can 
indeed be ignored, for the following reasons.  The two integrals 
stemming from Eqs.~(\ref{8.3a}) and (\ref{8.3c}) are formally independent of 
$a$, in accordance with our intuition that the self-energies of 
the walls are irrelevant to the force.  The contribution from the 
first term in Eq. (\ref{8.3b}) appears to be proportional to $a$, but 
again, we know from Refs.~\cite{hs,murray} that this term is precisely the 
zero-point energy inside the gap.

Even though we do not have an explicit expression for the fundamental
solution $F$, the first integral in Eq.~(\ref{U}) can be evaluated just
from the differential equation satisfied by $F$, as shown in Ref.~\cite{jw}:
\be
\int_0^\infty dz\, F^2(z)=\frac1{2\kappa}F(0)F'(0)\frac{d}{d\kappa}\ln\frac
{F(0)}{F'(0)}.
\ee
The latter may be readily expressed in terms of the reflection coefficient
$r$:
\be
\int_0^\infty dz\,F^2(z)=-\frac{F_-^2}{(2\kappa)^2}\left(\frac{r^2-1}{2\kappa}
+\frac{dr}{d\kappa}\right).
\ee
When this is substituted into Eq.~(\ref{U}), and terms independent of $a$ 
omitted, we obtain
\be
U\to -\frac1{12\pi^2}\int_0^\infty d\kappa\,\kappa^3\frac1{e^{2\kappa a}-r^2}
\left[-2r\frac{dr}{d\kappa}+2r^2a\right]=\frac1{4\pi^2}\int_0^\infty
d\kappa\,\kappa^2\ln\left(1-r^2e^{-2\kappa a}\right),
\ee
where the last step involves integration by parts. Evidently, differentiating
this with respect to $-a$ yields the pressure (\ref{lifshitz}), that is,
Eq.~(\ref{pvw}) is satisfied.  

As noted at the beginning, the Green's function is invariant under the
substitution $G\to G+p F$, where $p$ is independent of $z$. 
Here $p$ is not allowed to depend on the separation between the walls.
Such a substitution does not change the
Wronskian or the transmission coefficient, and changes the reflection
coefficient by a constant, $R\to R-p$.  Therefore, the energy (\ref{U})
changes only by a constant, and the Casimir pressure on one wall is unchanged.

\section{Conclusion}
We have in this paper significantly extended the analysis given in 
Ref.~\cite{hs}.  We now have extracted all the divergences corresponding
to the soft-wall potential, and have computed the energy density and stress
tensor within as well as outside the region of the potential,
for the case of linear and quadratic potentials.  The renormalized energy
density exhibits divergences as the boundary is approached, just as it does
in the case of a Dirichlet wall, but much weaker; these divergences
are the same (up
to a sign) on both sides of the wall.  The fact that the surface divergences
are proportional to $\xi-\frac16$ indicates the irrelevance of these terms,
since the total energy must be independent of the conformal parameter. 
However, before we can  ascribe a finite self-energy to this configuration, 
we must recognize that terms in the energy density
that grow with the distance into the wall require
physical interpretation.
It may be that the only physically unambiguously observable consequence
is the force between two soft walls, which we calculated in the last section
of this paper.

In future work, we hope to further understand the meaning of the energy
density, total energy, and stress in these configurations.  We hope to make
progress in solving the problem for general $\alpha$: In particular the limit
of $\alpha\to\infty$ would be of great interest to study, because that limit
would correspond to the appearance of a hard Dirichlet wall at $z=1$. 
(The emergence of this preferred length scale 
in a seemingly scale-invariant problem is related to the coupling 
constant that we have suppressed, as explained in Ref.~\cite{Bouas:2011ik}.)  
As $\alpha$ grows, the WKB approximation becomes increasingly unsuited 
to the region of small $\kappa$ and $z$, and hence it will be 
necessary to bring in approximations at small $\kappa$, completing 
the program of Ref.~\cite{murray}.  Accurate treatment of the contributions 
to the energy density from small $\kappa$ should clarify and remedy the 
deficiencies in the analysis offered in Appendix \ref{sec:surf}.

\acknowledgments
The work of KAM was supported in part by a grant from the Julian Schwinger
Foundation.  We thank Li Yang, Alex Mau, and Jacob Tice for collaborative
assistance, and Steve Christensen, Itay Griniasty, and Ulf Leonhardt 
for helpful conversations.

\appendix
\section{WKB approximation}
\label{sec:wkb}
In the text we are considering approximate solutions to the problem
\be y''(z)=Q(z)y(z),\quad Q(z)=\kappa^2+v(z),
\ee
with $Q$ positive and large.  The effective expansion parameter, 
therefore, multiplies $Q$ as a whole; one can write
$\epsilon^2 y'' = Qy$, $\epsilon\to0$.  (In quantum mechanics 
$\epsilon$ is identified with Planck's constant.)  We suppress 
$\epsilon$ (take it equal to~$1$) in the detailed formulas.
    
The WKB approximation is constructed to high order in \cite{bo} in 
terms of local functionals of $Q$, which we denote $q_n$, for 
each nonnegative integer~$n$:
\be      y(z) \sim \exp\left[\epsilon^{-1}\sum_{n=0}^\infty
(\pm) \int^z   \epsilon^n q_n(t) dt \right]. \ee
Fr\"oman \cite{froman} noted that the odd-order terms can be resummed 
into the prefactor:
\be
y(z)\sim \frac1{\sqrt{q_0(z)+q_2(z)+q_4(z)+\dots}}
e^{\pm \int^z dt[q_0(t)+q_2(t)+q_4(t)+\dots]},\label{genwkb}
\ee
where $q_n$ is accompanied by $\epsilon^{n-1}$ in the exponent 
and by $\epsilon^n$ in the prefactor.  The first three even-order 
WKB integrands are (in the notation of Bender and Orszag \cite{bo})
\begin{subequations}\label{wkbq}
\bea
q_0(t)&=&Q^{1/2}(t),\label{wkbq0}\\
q_2(t)&=&\frac{Q''(t)}{8Q^{3/2}(t)}-\frac5{32}\frac{Q^{\prime2}}{Q^{5/2}(t)},
\label{wkbq2}\\
q_4(t)&=&\frac{Q^{(4)}(t)}{32 Q^{5/2}(t)}-\frac7{32}\frac{Q'(t)Q'''(t)}{Q^{7/2}
(t)}-\frac{19}{128}\frac{Q^{\prime\prime2}(t)}{Q^{7/2}(t)}+\frac{221}{256}
\frac{Q''(t)Q^{\prime2}(t)}{Q^{9/2}(t)}-\frac{1105}{2048}\frac{Q^{\prime4}(t)}
{Q^{11/2}(t)}.
\eea
\end{subequations}
(In Ref.~\cite{froman}, $Q$ is called $Q^2$.)  
In the Fr\"oman approximation of 
order $2n$, the exponential sum terminates with $q_{2n}$ and the 
prefactor sum terminates with $q_{2n-2}$.  In the approximation 
of order $2n+1$, both series extend through $q_{2n}$.

When these successive WKB approximants are used in computing the first term
in the diagonal Green's function (\ref{gee}), orders $2n$ and $2n+1$ give 
identical results for the particular combination that is relevant,
\be
\frac{F(z)G(z)}w=\left(\frac{G'(z)}{G(z)}-\frac{F'(z)}{F(z)}\right)^{-1}
\sim\frac12\frac1{q_0+q_2+q_4+\dots}.
\ee
(Recall that $F$ is the solution which vanishes exponentially at 
positive infinity, so $G$ must be dominated by the exponentially growing 
solution.)   Continuing to expand 
in powers of $\epsilon$:           
\be
\frac{FG}{w}(z)\sim \frac1{2q_0(z)}\left(1-\frac{q_2(z)}{q_0(z)}+
\left[\left(\frac{q_2(z)}{q_0(z)}\right)^2-\frac{q_4(z)}{q_0(z)}\right]+
O(\epsilon^6)\right).
\label{wkbexp}
\ee
The zeroth order WKB term yields the first term displayed in Eq.~(\ref{2wkb}),
while the 2nd and 3rd terms there result from the second-order term in 
Eq.~(\ref{wkbexp}).  The two terms in the square brackets in Eq.~(\ref{wkbexp})
give the 4th-order contribution to the coincident Green's function,
\be
g^{(4)}(z,z)=\frac1{2\sqrt{Q}}\left(-\frac1{32}\frac{v^{(4)}}{Q^3}+\frac7{32}
\frac{v'v'''}{Q^4}+\frac{21}{128}\frac{v^{\prime\prime2}}{Q^4}-\frac{231}{256}
\frac{v''v^{\prime2}}{Q^5}+\frac{1155}{2048}\frac{v^{\prime4}}{Q^6}\right).
\ee
Note that an expansion in $\epsilon$ is not quite the same 
thing as one in $1/\kappa$.  Including enough WKB terms is 
sufficient but not necessary to obtain a certain order in 
$\kappa$.  Thus in Eq.~(\ref{2wkb}) the second term was necessary to 
capture all divergences \cite{hs}, whereas the third term was not but 
is needed to capture the correct WKB behavior at large~$z$.

To compute all the components of the stress tensor, we have to expand $I[g]$
out to order $\delta^2$.  The corresponding terms are
\begin{subequations}
\bea
4\pi^2I[g^{(0)}]&=&\frac1{\delta^2}+\frac{v}2\left(\ln\frac{\sqrt{v}\delta}2+
\gamma-\frac12\right)
+\frac{v^2\delta^2}{16}\left(\ln\frac{\sqrt{v}\delta}2
+\gamma-\frac54\right),\label{I0}\\
4\pi^2 I[g^{(2)}]&=&\frac1{24}\left(-\frac{v''}v+\frac12\frac{v^{\prime2}}{v^2}
\right)-\frac{v''\delta^2}{48}\left(\ln\frac{\sqrt{v}\delta}2+
\gamma-\frac12\right)-\frac{\delta^2}{192}\frac{v^{\prime2}}{v},\label{I2}\\
4\pi^2 I[g^{(4)}]
&=&-\frac1{240}\frac{v^{(4)}}{v^2}
+\frac1{60}\frac{v'v''}{v^3}+\frac1{80}
\frac{v^{\prime\prime2}}{v^3}-\frac{11}{240}\frac{v^{\prime2}v''}
{v^4}+\frac1{48}\frac{v^{\prime4}}{v^5}\nn\\
&&\quad\mbox{}+\delta^2\left[\frac1{960}\frac{v^{(4)}}v-\frac1{480}
\frac{v'v'''}{v^2}-\frac1{640}\frac{v^{\prime\prime2}}{v^2}
+\frac{11}{2880}\frac{v''v^{\prime2}}{v^3}-\frac1{768}\frac{v^{\prime4}}{v^4}
\right].\label{I4}
\eea
\end{subequations}

\section{``Surface'' divergence}
\label{sec:surf}
Here we examine the behavior of the energy as the boundary at $z=0$ is
approached from above.  We confine attention 
to the cases of interest in this paper, $\alpha=1$ and $2$, and 
to the contribution from the WKB region of the Euclideanized 
spectrum, which appears to be the most important one.

Recall that the Green's function has the construction [see Eqs.~(\ref{gee})--%
(\ref{wronskian}) and (\ref{abbrev})]
\begin{subequations}
\bea
z,z'>0:\quad g(z,z')
&=&\frac1w F(z_>)G(z_<)-\frac{G_-}{F_-}\frac1w F(z)F(z'),\label{ginwall}\\ 
z,z'<0:\quad g(z,z')
&=&\frac1{2\kappa} e^{-\kappa(z_>-z_<)}+\frac{F_+}{F_-}\frac1{2\kappa} 
e^{\kappa(z+z')}.\label{gbelow}
\eea
\end{subequations}
When $z'=z>0$, we introduce a short notation for the two terms in 
Eq.~(\ref{ginwall}),
\be g_{FG} = \frac1w F(z)G(z), \qquad g_{FF} =
 -\,\frac1w\,\frac{G_-}{F_-}\,F(z)^2.   
\label{new}
\ee
In the case $\alpha=1$ (the linear wall), the exponentially 
decreasing solution~$F$ 
can be chosen as $\Ai(\kappa^2+z)$, and the independent solution
$G$ can be chosen as $\Bi(\kappa^2+z)$.

In Secs.~\ref{sec:asy} and \ref{sec:lin} 
we used the WKB approximation only on the first term in Eq.~(\ref{ginwall}). 
 As we have seen, although it captures the 
correct behavior for large $\kappa$, this procedure generates 
spurious singularities for $z$ near the boundary, presumably 
stemming from the inadequacy of the WKB approximation at 
small~$\kappa$ and the neglect of the second term, $g_{FF}$. 
We argue that this first term is, in fact, not relevant to the 
question of ``surface divergences''.  
The corresponding energy density is given by
\be
u_{FG}=\left(\frac{\partial^2}{\partial\tau^2}-\beta\frac{\partial^2}{\partial 
z^2}\right) I[g_{FG}].
\ee
For $\alpha=1$, the explicit form of $g_{FG}$ appears as the first term in 
Eq.~(\ref{lineargee}). Use of 
the asymptotic expansions of the Airy functions for large 
argument gives, of course, the WKB result (\ref{wkb}).  
But suppose, on the 
contrary, that we simply subtract (even at positive $z$) the 
first term in Eq.~(\ref{gbelow}), which would produce the free-field 
zero-point energy.  That is, we replace $g_{FG}$ by $g_{FG}-1/(2\kappa)$.
Then one can easily check numerically that 
$I[g_{FG}-1/(2\kappa)]$  has a finite second derivative with 
respect to $\tau$ for $z\to0$ and a finite second derivative with 
respect to $z$ at $z=0$.
Thus, as expected, no surface divergence originates from this 
term.
(The modifications introduced by renormalization are nonsingular 
at $z=0$.)

On the other hand, the  WKB expansion 
is effective for
isolating the small-$z$ behavior of the energy arising from $g_{FF}$.
This expansion is valid for large $\kappa$, even for small~$z$.
The asymptotic behaviors of the Airy functions are \cite{bo}
\begin{subequations}
\bea
\mbox{Ai}(z)&\sim& \frac1{2\sqrt{\pi}}z^{-1/4}e^{-2z^{3/2}/3}
\left(1-c_1 z^{-3/2}+O(z^{-3})\right),\\
\mbox{Bi}(z)&\sim& \frac1{\sqrt{\pi}}z^{-1/4}e^{2z^{3/2}/3}
\left(1+c_1 z^{-3/2}+O(z^{-3})\right)+O\left(e^{-2 z^{3/2}/3}\right),
\label{bia}
\eea
\end{subequations}
where $c_1=5/48$.  Extrapolating to $z=0$ (and dropping some 
$\kappa$-independent constants), these formulas 
suggest the initial data
\begin{subequations}
\label{genas}
\bea
F(0)&\sim&\frac1{\sqrt{\kappa}}e^{-\Lambda(\kappa)},\quad 
F'(0)\sim-\sqrt{\kappa}\left(1+\frac{v'(0)}{4\kappa^3}\right)
e^{-\Lambda(\kappa)},\\
G(0)&\sim&\frac1{\sqrt{\kappa}}e^{\Lambda(\kappa)},\quad
G'(0)\sim \sqrt{\kappa}\left(1-\frac{v'(0)}{4\kappa^3}\right)
e^{\Lambda(\kappa)}\label{gag},
\eea
\end{subequations}
where $\Lambda(\kappa)=\frac23\kappa^{3/2}$.  
We have written Eqs.~(\ref{genas}) in  a form that identifies them with
the first-order WKB formulas (\ref{genwkb}) and (\ref{wkbq0}) 
for a particular choice of normalization, which makes the Wronskian 
independent of $\kappa$ ($w=2$, to be precise).    This normalization can 
usefully be copied for dealing with other values of $\alpha$.
Note that $\Lambda \to \infty$ as $\kappa\to\infty$.
Validity of Eq. (\ref{gag}) requires that $\kappa$ be sufficiently 
large that both (a) the WKB approximation is accurate and (b)
the recessive term in Eq. (\ref{bia}) is negligible.
From Eqs.~(\ref{genas}) it follows that
\be
\frac{G_-}{F_-}=
\frac{\kappa \mbox{Bi}(\kappa^2)-\mbox{Bi}'(\kappa^2)}
{\kappa \mbox{Ai}(\kappa^2)-\mbox{Ai}'(\kappa^2)}\sim \frac1{4\kappa^3}e^{4
\kappa^3/3},\label{largekapparc}
\ee
($c_1$ having cancelled), and hence
\be
g_{FF}\sim
-\frac1{16}\frac{v'(0)}{\kappa^4}e^{-2\kappa z}.\label{f2piece}
\ee
The resulting term in $u$ diverges at the boundary,
\be
\alpha=1:\quad u_{\rm surf}=\frac{1+12\beta}{192\pi^2z},\label{us1}
\ee
as reported in Eq.~(\ref{usurf}) and numerically validated in 
Fig.~\ref{ulinint} . (This calculation extrapolates the integrand 
(\ref{f2piece}) 
down to $\kappa=0$.  In principle we know how to improve it by the method 
of Ref.~\cite{murray}.)

It is now incumbent upon us to investigate in what way this 
result is dependent upon the ``handbook'' basis choice,
$\{\Ai,\Bi\}$.
As we have stressed repeatedly (Refs.~\cite{hs,murray}, and the body of this 
paper),  the Green's function must not change under rescalings 
$F\to mF$, $G\to nG$, nor under a replacement $G\to G + pF$,
where $m$, $n$, and $p$ may depend on $\kappa$.  The rescalings
are trivially taken care of by the Wronskian factors $w$ in Eq.~(\ref{new}),
so long as one has resisted the temptation to replace $w$
by its value in some particular basis.  The $p$ replacement is 
more subtle, however; although $g$ of course remains unchanged, 
its division into the two terms of Eq.~(\ref{new}) does not.  In 
particular, one might choose $G$ so that $G_-$ is identically 
$0$.  (This is the case for the solution called $H$ in Ref.~\cite{murray}.)
Then $g_{FF}=0$, and $u$ must come entirely from $g_{FG}$.
Thus our attribution of the surface energy to the second term of 
the Green's function cannot be valid in complete generality.
What is going on here?  Let us return to Eq.~(\ref{largekapparc}) and consider 
replacing $\Bi$ by $\Bi + p\Ai$.  One sees that any 
admixture of $\Ai$ will give an exponentially subdominant 
contribution, unless $p(\kappa)$ contains a correspondingly 
large exponential factor.  This suggests that our calculation
captures the truth  for any ``natural'' 
basis choice, one not involving such an exponential fine tuning.
One can easily see that the preferred solution called $G$
in Ref.~\cite{murray} (characterized by $G(0)=0$) is proportional to
$\Bi + p \Ai$ with $p = -2 e^{4\kappa^3/3}(1 + 2 c_1 \kappa^{-3} 
+\cdots)$.  For $H$, the other preferred solution in Ref.~\cite{murray}, the 
calculation is more complicated, but again $p$ will equal
$e^{4\kappa^3/3}$ times a weakly (algebraically) varying function 
of $\kappa$.   Such basis solutions, however natural for our 
problem, must be regarded as rare.

Note that a small admixture of $\Ai$ will not change the 
initial data (\ref{gag}) significantly.  
In fact, there is no reason to expect that $\Bi$ satisfies
Eq. (\ref{gag}) or any higher-order improvement of it exactly
[equivalently, that the recessive term in (\ref{bia}) is exactly zero].
It is therefore legitimate to challenge the numerical 
verification that  $g_{FG}$ yields no surface divergence at all;
more likely, one is present but with such a tiny coefficient that
it did not show up in the  finite-precision numerical 
investigation.

Can these considerations be carried over to larger $\alpha$?
Since then $v'(0)=0$, Eqs.~(\ref{genas}) lead to trivial results and must 
be replaced by higher-order approximations.  For the quadratic 
wall we go out to third WKB order [in the sense of Appendix A ---
that is, keeping $q_2$ in both the exponent and the prefactor of
Eq.~(\ref{genwkb})] and obtain from Eqs.\ (\ref{genwkb})--(\ref{wkbq2})
\begin{subequations}
\bea
F(0)&\sim&
\left(\kappa+\frac1{4\kappa^3}\right)^{-1/2}e^{-\Lambda(\kappa)},\quad
F'(0)\sim-\left(\kappa+\frac1{4\kappa^3}\right)^{1/2}e^{-\Lambda(\kappa)},
\label{fqa}\\
G(0)&\sim&\left(\kappa+\frac1{4\kappa^3}\right)^{-1/2}e^{\Lambda(\kappa)},\quad
G'(0)\sim\left(\kappa+\frac1{4\kappa^3}\right)^{1/2}e^{\Lambda(\kappa)}.
\label{fqb}
\eea
\end{subequations}
Here Eq.~(\ref{fqa}) is ineluctable, but Eq.~(\ref{fqb}) incorporates the
tacit assumption that $G$ is a ``natural'' basis solution without 
a large recessive component, so that the surface divergence will 
come entirely from $g_{FF}$.
[If the recessive term in $G$ is significant at large $\kappa$, 
then (a) it may make a surface-divergent contribution to the 
$g_{FG}$ term and (b) it may cause a compensating change in the 
$g_{FF}$ term through the factor $G_-$.  These two effects must 
cancel when the two terms are known exactly, since the full 
answer must be independent of the basis choice.]
Now for large $\kappa$ 
\be
F(0)-\frac1\kappa F'(0)\sim\frac2{\sqrt{\kappa}}e^{-\Lambda(\kappa)},\quad
G(0)-\frac1\kappa G'(0)\sim-\frac14\frac{1}{\kappa^{9/2}}e^{\Lambda(\kappa)},
\ee
and hence
\be
g_{FF}\sim
\frac1{16}\frac1{\kappa^5}e^{-2\kappa z}. \label{gf2q}
\ee
In this case the resulting integral for $u$ diverges at the lower 
limit, $\kappa=0$, so as in Ref.~\cite{hs} it must be cut off at, say,
$\kappa=1$, yielding
\be
\alpha=2:\quad u_{\rm surf}=-\frac{1+12\beta}{96\pi^2}\Gamma(0,2z).\label{us2}
\ee
The ambiguity in this infrared cutoff 
(which would  not be necessary at all in a more accurate 
treatment of small $\kappa$ \cite{murray})
can be absorbed into the 
logarithmic ambiguity $\mu$ from the (ultraviolet) renormalization.
  As reported in Sec.~\ref{sec:quad}, 
after this one undetermined constant is fixed, Eq.~(\ref{us2}) agrees with the 
numerics.
[The leading asymptotic correction to $g_{FF}$ for $\alpha=2$
was used to approximate the integral from $K$ to $\infty$ in
Eqs.~(\ref{twoparts}) and (\ref{largeK}), 
so that part of the numerical agreement 
was foreordained.  However, that part of the integral is only a 
small part of the total, and the numerical confirmation of 
Eq.~(\ref{quadfit}) is nontrivial.]

Furthermore,  both  Eq. (\ref{us1}) and Eq. (\ref{us2}) match the calculations 
in Ref.~\cite{hs} for the exterior region, strengthening our confidence 
that the nonrigorous argument in this appendix reflects reality.  
More precisely,  result (\ref{us2}) is exactly the negative of the 
density found on the other side of the wall in Ref.~\cite{hs}, with
$z\to|z|$, while result (\ref{us1}) is the same as in Ref.~\cite{hs}, 
including the sign.

The argument shows that there is no surface divergence for $T_{zz}$, because
in a term  proportional to $e^{-2\kappa z}$, 
in the construction (\ref{tzz}) the leading powers of $\kappa$
cancel and leave a positive power of $z$:
\be
T_{zz}\sim -z^\alpha I[g_{F^2}].
\ee

\end{document}